\documentclass[]{tPHM2e}

\begin{document}
\doi{10.1080/1478643YYxxxxxxxx}
\issn{1478-6443}
\issnp{1478-6435}
\jvol{00} \jnum{00} \jyear{2008} \jmonth{21 December}

\markboth{Jop, Petrosyan and Ciliberto}{Philosophical Magazine}

\articletype{Special Issue}

\title{{\itshape Effective Temperature in a Colloidal Glass}}

\author{Pierre Jop , Artyom Petrosyan and Sergio Ciliberto$^{\ast}
$\thanks{$^\ast$Corresponding author.
Email:sergio.ciliberto@ens-lyon.fr \vspace{6pt}}\\\vspace{6pt}
{\em{Universit\'e de Lyon, Laboratoire de Physique,\\ Ecole
Normale Sup\'erieure de  Lyon, CNRS ,\\
        46, All\'ee d'Italie, 69364 Lyon CEDEX 07, France}}\\\vspace{6pt}\received{Mai 2008} }

\maketitle

\begin{abstract}
 We study the Brownian motion of particles trapped by optical tweezers
 inside a colloidal glass (Laponite) during the
 sol-gel transition. We use two methods based on  passive
 rheology to extract the effective temperature from the fluctuations of the Brownian particles.
 All of them give a temperature that, within experimental errors, is equal to the heat bath
 temperature. Several interesting features concerning the statistical properties
 and the long time correlations of the particles are observed during the transition.

 \bigskip

%\centerline{\bfseries Index }\medskip

%\hbox to \textwidth{\hsize\textwidth\vbox{\hsize18pc
%\hspace*{-12pt} {1.}    Introduction\\
%{2.}    Experimental set-up\\
%\hspace*{10pt}{2.1.}  Laser modulation technique \\
%\hspace*{10pt}{3.2.}  Active rheology\\
%\hspace*{10pt}{2.2.}  Kramers-Kronig and modulation technique\\
%{3.}    Experimental results\\
%\hspace*{10pt}{3.1.}  Laser modulation method \\
%\hspace*{10pt}{3.2.}  Active rheology\\
%\hspace*{10pt}{3.2.}  Passive rheology\\
%{4.}   Results using multiple traps \\
%\hspace*{10pt}{4.1.}   Long range correlations\\
%\hspace*{10pt}{4.2.}   Passive rheology with two traps\\
%{4.}  Conclusion}}
\end{abstract}

\section{Introduction}

The glasses and colloids are interesting examples of out of
equilibrium systems where the relaxation towards the equilibrium
may last much more than a reasonable observation time. One of the
problem which has been widely theoretically  studied is the
definition of an effective temperature $T_{eff}$ in these systems
using the fluctuation dissipation relation (FDR). This relation is
an extension of the Fluctuation Dissipation Theorem for an out of
equilibrium system and $T_{eff}$ is defined as the ratio between
the correlation and the response function\cite{Cugliandolo}. Using
numerical simulations  it has been shown that in several models
for out of equilibrium systems the effective temperature defined
via FDR  is higher than the temperature of the thermal bath and it
is a good definition of temperature in the thermodynamic sense.
However the experimental results are more confused. It has been
observed that $T_{eff}$ in polymer, spin glasses and colloids may
depend on the experimental conditions. For example in the
dielectric measurements in polymer $T_{eff}$ depends on the
quenching rate and it may be huge because of the presence of
intermittent bursts. The same behavior is observed on mechanical
variables. Instead in a colloid (Laponite) during the sol-gel
transition the electric observables give a $T_{eff}$ which is
quite large whereas measurements done on thermal rheometer
indicate that within experimental errors there is no violation of
the Fluctuation theorem because the $T_{eff}$ is always equal to
the temperature of the thermal bath.(A discussion on the $T_{eff}$
obtained from dielectric measurements and mechanical measurements
can be found in  ref.\cite{Bellon02}). The rheological measurement
described in ref.\cite{Bellon02} was a global measurement and one
may wonder whether an experiment of microrheology give or not the
same results. This experiment can be performed using as a probe a
Brownian particle using active and passive microrheology. These
kind of experiments are interesting also from another point of
view because one may studies whether the properties of the
Brownian motion are affected by the fact that the surrounding
fluid (the thermal bath) is out of equilibrium. Several
experiments of Brownian motion inside a Laponite solution have
been done by different groups using various techniques. The
results are rather contradictory. Let us resume them. Abou et al.
find an increase and then a decrease of $T_{eff}$ as a function of
time \cite{Abou06}. Bartelett et al. find an increase of $T_{eff}$
\cite{Greinert07}. In contrast Jabbari-Farouji et al. do not
observe any change and confirm the results on the thermal
rheometer of ref \cite{Bellon02}. The purpose of this article is
to describe the results of the measurement of the Brownian motion
of a particle inside a Laponite solution using a
 combination of  different techniques proposed in previous references.
 All the techniques do not show, within experimental errors, any
 increase of $T_{eff}$ which  remains equal to that of the
 thermal bath for all the duration of the sol-gel transition.
Thus the result of this papers agrees with those of
Jabbari-Farouji et al. \cite{Jabbari-Farouji07} and of Bellon et
al. \cite{Bellon02}.
 The article is organized as follow. In section 2 we describe the
 experimental apparatus and the various techniques used to measure
 response and fluctuations. In section 3 we describe the
 experimental results of the various experimental techniques.
 In section 4 we discuss the result and we conclude.

\section{Experimental set up}

We measure the fluctuations of the position of one or several
silica beads trapped by an optical tweezers during the aging of
the Laponite. The laser beam ($\lambda$=980 nm) is focused by a
microscope objective ($\times$63) 20 $\mu$m above
 the cover-slip surface to create a
 harmonic potential well where a bead of 1 or 2 $\mu$m in diameter ($2r$) is trapped.
 We can  trap several beads  if the laser is rapidly swept from a position to another
 to form a multi trap system.
The Laponite mass concentration is varied from 1.2 to 3\% wt for
different ionic strengths. These conditions allow us to obtain
either a gel or a glass according to the phase diagramm found in
the literature \cite{bonn_PRE_2004}. The Laponite is filtered with
a 0.45 $\mu$m filter
 to avoid the formation of aggregates. The aging time $t_w$
is measured since the end of this filtering process. Particular
attention has been
 paid in the cell construction
 indeed the properties of the Laponite are very sensitive
 to the experimental conditions. First the Laponite solution
 is prepared under nitrogen atmosphere and second the cell
 is completely sealed using
 Gene Frame adhesive spacers in order to avoid evaporation and
 the use of vacuum grease, used in other experiments. Indeed this grease, whose
  PH is much smaller than 10, quickens the evolution of the suspension.
%......

In order to measure $T_{eff}$ for the particle motion two
techniques have been used. The first one is based on the laser
modulation technique as proposed in ref.\cite{Greinert07}. The
second is based on the Kramers-Kronig relations with two laser
intensities and it combines the advantages of two methods one
proposed in ref.\cite{Jabbari-Farouji07} and in
ref.\cite{Greinert07}.

\subsection{Laser modulation technique}

Following the method used in \cite{Greinert07}, the stiffness of
the optical trap is periodically switched between two different
values ($k_1=6.34$ pN/$\mu$m and $k_2=14.4$ pN/$\mu$m) every 61
seconds by changing the laser intensity. The position of the bead
is recorded by a quadrant photodiode at the rate of 8192 Hz, then
we compute the variance of the position over the whole signal,
$\left<\delta x^2\right>=\left< x^2-\left<x\right>^2\right>$,
where the brackets stand for average over the time. To avoid
transients, each record is started 20 seconds after the laser
switch to be sure that the system has relaxed toward a quasi
stationary state.
 Assuming the equipartition principle still holds
  in this out-of-equilibrium system, the $T_{eff}$
 is computed as in \cite{Greinert07}
 the expression of the effective temperature
 and of the Laponite elastic stiffness $K_{Lap}$
  is the following:
\begin{equation}
   k_BT_{eff}={
   (k_2 - k_1)\left<\delta x_1^2\right>\left<\delta x_2^2\right>
   \over \left<\delta x_1^2\right>-\left<\delta x_2^2\right> }
\label{eq:temperature}
\end{equation}
\begin{equation}
   K_{Lap}=
   {k_1\ \left<\delta x_1^2\right>- k_2\left<\delta x_2^2\right>
   \over
   \left<\delta x_1^2\right>-\left<\delta x_2^2\right>}
\label{eq:KLap}
\end{equation}

This technique, although quite interesting and simple,  has the
important drawback that, being a global measurement, it has no
control of what is going on the different frequencies. To overcome
this problem we have used the following method.

\subsection{Kramers-Kronig and modulation technique}

This combines  the laser modulation technique described in the
previous section and the passive rheology technique based on
Kramers-Kronig relations. The fluctuation dissipation relations
relate the spectrum $S_i(\omega)$ of the fluctuation of the
particle position to the imaginary part $\alpha_i''$ of the
response of the particle to an external force, specifically :

\begin{equation}
S_i(\omega,t_w)= {4 \ k_B T_{eff} \over \omega}
\alpha_i''(\omega,t_w) \label{eq:FDR}
\end{equation}

with $i=1$ and $i=2$ for the spectra measured with the trap
stiffness $k_1$ and $k_2$  respectively. We recall that for a
particle inside a  newtonian fluid $G=1/\alpha_i= k_i+ i\nu
\omega$ with $\nu=6\pi\eta r$ and $\eta$ the fluid viscosity. In
Eq. \ref{eq:FDR} the dependence in $t_w$ takes into account the
fact that the properties of the fluid changes after the
preparation of the Laponite. If one assumes that $T_{eff}$ is
constant as a function of frequency (hypothesis that can be easily
checked a posteriori) then the real part $\alpha_i'$ of the
response is related to $\alpha_i''$ by the Kramers-Kronig
relations \cite{KK_R} that is:
\begin{equation}
\alpha_i'(\omega,t_w)={ 2\over \pi } P\int_0^\infty {\xi
\alpha''(\xi,t_w)  \over \xi^2-\omega^2} d\xi ={ 1\over 2 \pi k_B
T } P\int_0^\infty {\xi^2 S_i(\xi,t_w)  \over \xi^2-\omega^2} d\xi
\label{eq:KrKr}
\end{equation}
where $P$ stands for principal part of the integral and assumed
that $T_{eff}=T$ to write the second equality using
Eq. \ref{eq:FDR}. However it can be easily (details will be given
in a longer report) shown that using the two measurements at $k_1$
and $k_2$ we get:
\begin{equation}
T_{eff}(\omega,t_w)= T_{bath} \left({k_1- k_2 \over
G'_1(\omega,t_w)- G'_2(\omega,t_w) }\right)
\end{equation}
and $G_i'(\omega,t_w)=k_i+K_{Lap}(\omega,t_w)$.  It is clear that
if one finds  a dependence of $T_{eff}$ on $\omega$ this method
cannot be used because  $\alpha''(\omega,t_w)$ is not simply
proportional to $S_i(\omega,t_w) \omega$ as assumed in
Eq. \ref{eq:KrKr}.

\section{Experimental results}

\subsection{Laser modulation method}

\begin{figure}
\begin{center}
\subfigure[]{
\resizebox*{6cm}{!}{\includegraphics{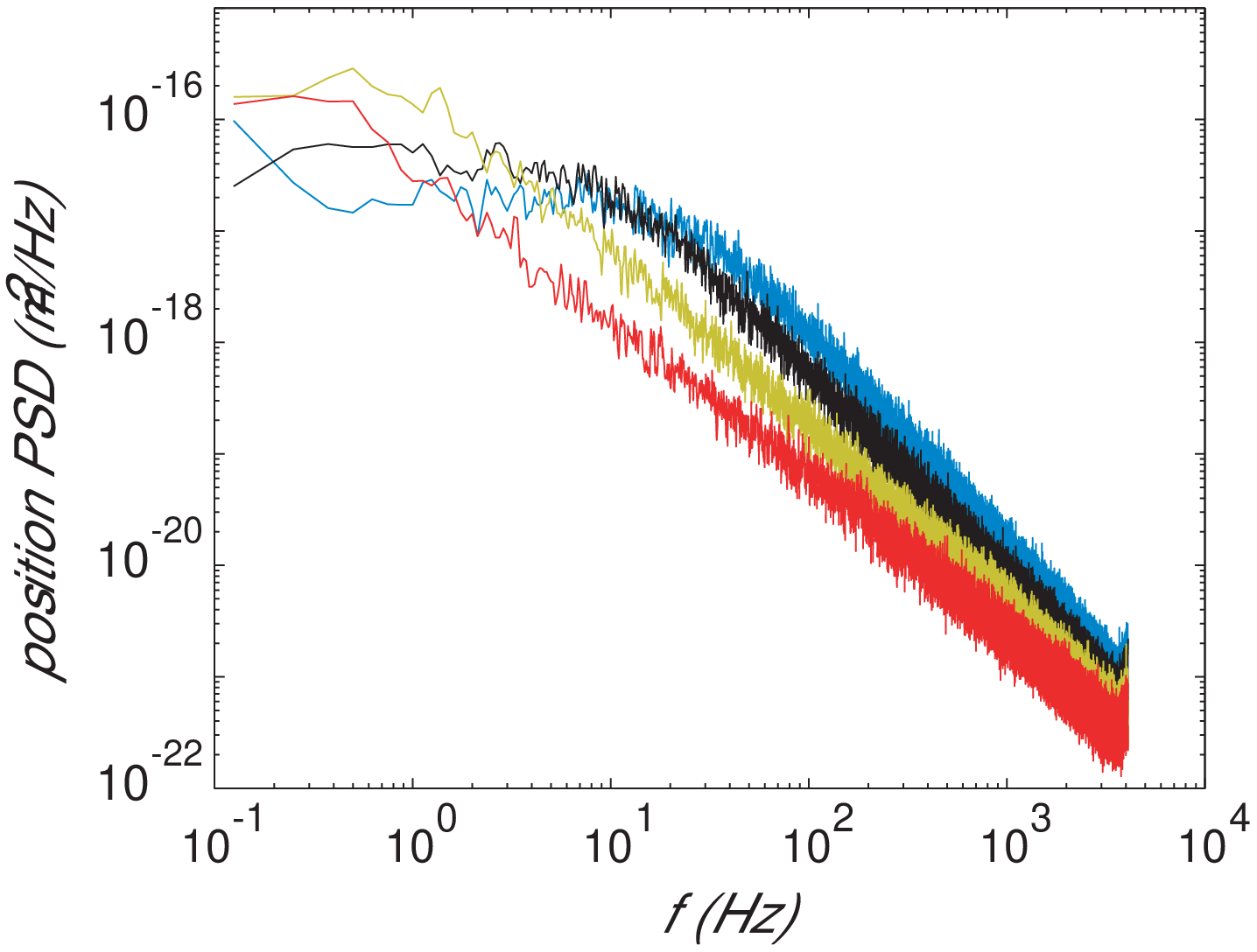}}}
\subfigure[]{
\resizebox*{6cm}{!}{\includegraphics{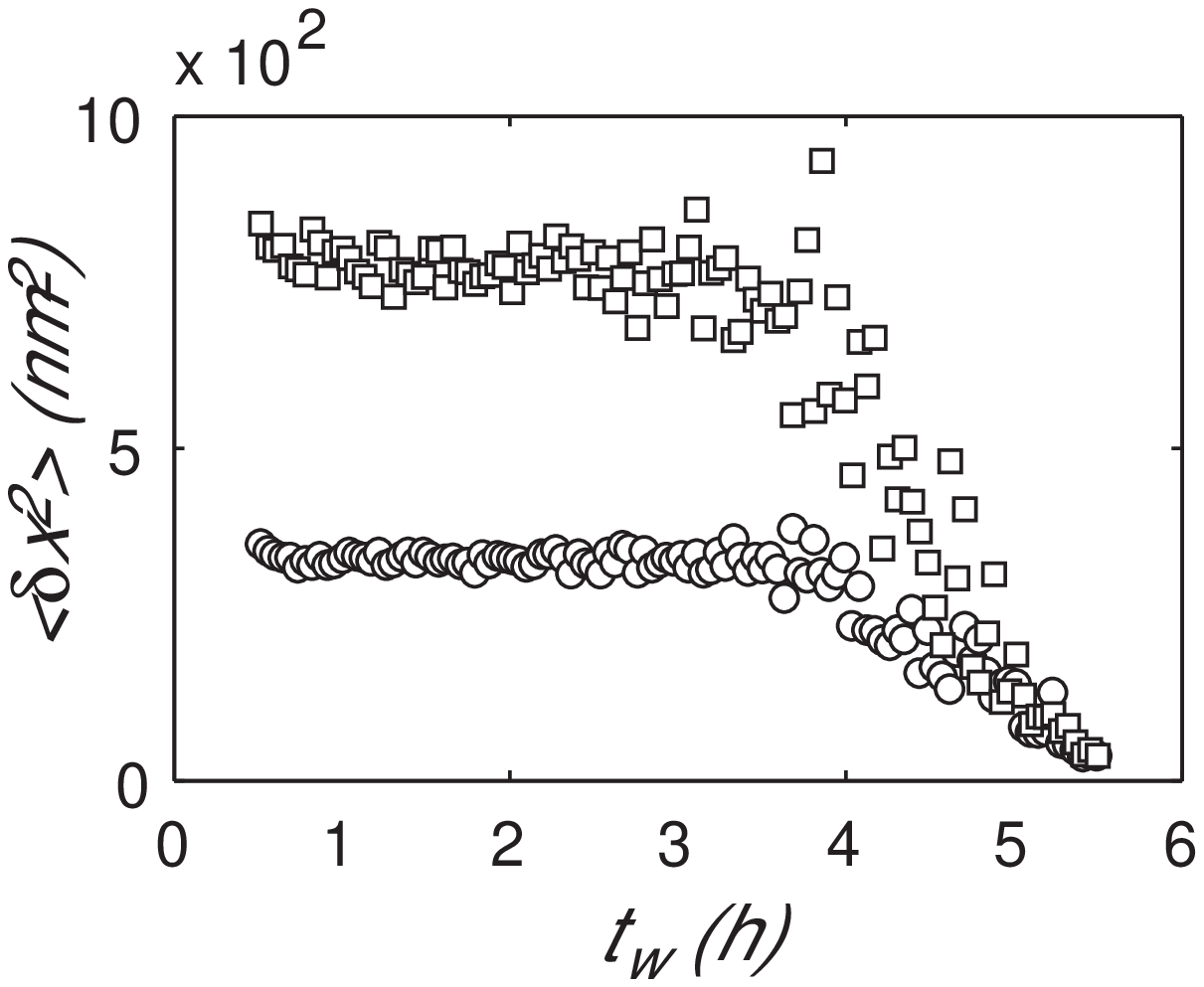}}}
\caption{\label{fig:spectra} a) Spectra of the particle position
fluctuations inside a Laponite solution at $2.3\%$ wt. The spectra
has been measured for the trap stiffness $k_1=6.34$ pN/$\mu$m
 at various ageing times: from right to left
$t_w=30$,  $110$, $190$ and $270$ minutes. b) Evolution of
$\left<\delta x^2\right>$ with $t_w$ for both stiffness,
$k_1=6.34$ pN/$\mu$m  ($\circ$) and $k_2=14.4$ pN/$\mu$m,
($\square$). The variances have been measured on a time window of
$\tau= 61$s}
\end{center}
\end{figure}

\begin{figure}
\begin{center}
%%\subfigure[]{
{\resizebox*{13cm}{!}{\includegraphics{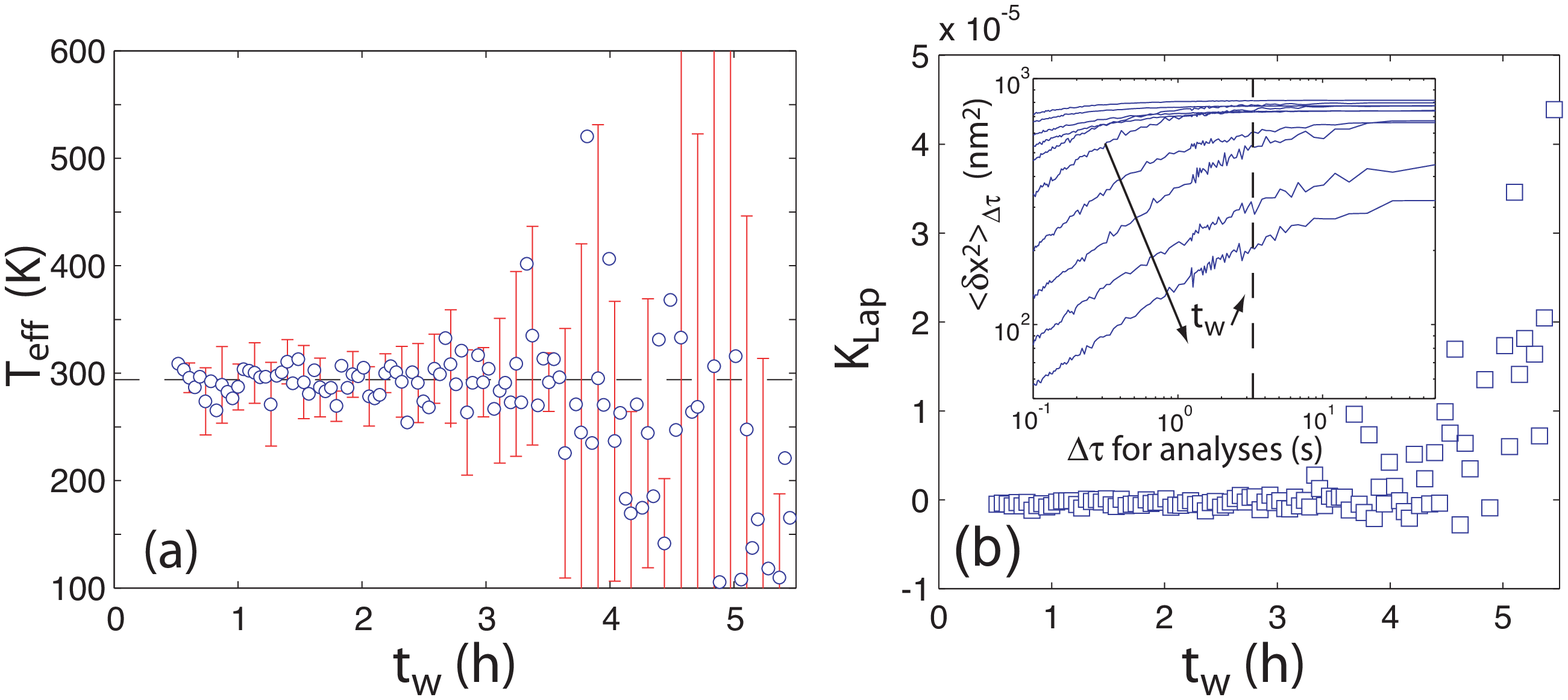}}}
%{\resizebox*{10cm}{!}{\includegraphics{errbetvarian5.eps}}\\
%\includegraphics[width=6cm]{stiffness-modul.eps}}
\caption{\label{fig:Teff} a) Evolution of the effective
temperature with aging time for the data of  Fig. \ref{fig:spectra} $k_1=6.34$ pN/$\mu$m, $k_2=14.4$ pN/$\mu$m, 2.3\%
 wt of Laponite and 1 $\mu$m glass bead. The error bars are computed from the statistical error of $\left<\delta x_i^2\right>$ at the time $t_w$.
  b) Evolution of the stiffness of the Laponite with ageing time.
  Inset: Evolution of $\left<\delta x^2\right>_{\Delta \tau}$ for $k_1$ as a function of the duration $\Delta \tau$
  of the samples for different aging times : $t_w=30$, $57$, $83$, $110$, $137$, $163$, $190$, $217$, $243$ and $270$ minutes.}
\end{center}
\end{figure}

Fig. \ref{fig:spectra}a) shows the power spectra of the particle
fluctuations inside Laponite at concentration measured at four
different $t_w$ with the trap stiffness $k_1=6.34$ pN/$\mu$m. We see
that  as time goes on the low frequency component of the spectrum
increases. That is the frequency cut-off $(k_i+K_{Lap})/\nu$
decreases mainly because of the increasing of the viscosity.  At
very long time this cut-off is well below $0.1Hz$.
In Fig. \ref{fig:spectra}b) we plot the variance of the
particle measured for the same data of Fig. \ref{fig:spectra}a) on
time windows of length $\tau=61$ s. The variance  remains constant
for a very long time and they begin to decrease because of the
increase o the gel stiffness. Using these data and
Eq. \ref{eq:temperature} and Eq. \ref{eq:KLap} one can compute
$T_{eff}$ and $K_{Lap}$. The results for $T_{eff}$ and $K_{Lap}$
are shown in Fig. \ref{fig:Teff}a) and  \ref{fig:Teff}b) respectively.

 We find that $T_{eff}$ is constant at the beginning and is very close to $T_{bath}=294$ K,
 then when the jamming occurs, that is when $K_{Lap}$ increases,
 it becomes more scattered without any clear increase with $t_w$, contrary to Ref.
\cite{Greinert07}. We now make several remarks. First, we point
out that the uncertainty of their results are underestimated. The
error bars in Fig. \ref{fig:Teff}a) are here evaluated from the
standard deviation of the variance using Eq.~2 in Ref
\cite{Greinert07} at the time $t_w$. Although they are small for
short time $t_w$, ($\Delta  T_{eff}/T_{eff}\leq 10\%$), they
increase for large $t_w$. This is a consequence of the increase of
variabilities of $\left<\delta x_i^2\right>$ as the colloidal
glass forms. This point is not discussed in in Ref.
\cite{Greinert07} and we think that the measurement errors are of
the same order or larger than the observed effect. The results
depend on the length of the analyzing time window and the use of
the principle of energy equipartition becomes questionable for the
following reasons.

%autre thechnique  <- melange de bartlett+Jabbari
First, these analyzing windows cannot be made too large because
the viscoelastic properties of Laponite evolve as a function of
time.
 Second, the corner frequency of the global trap (optical trap and gel),
 the ratio of the trap stiffness to viscosity,
 decreases continuously mainly because of the increase of viscosity.
 At the end of the experiment,
 the power spectrum density of the displacement of the bead shows that
 the corner frequency is lower than $0.1$ Hz.
We thus observe long lived fluctuations, which could not be taken
into account with short measuring times. This problem is shown on
Fig.~\ref{fig:Teff}b). We split our data into equal time
duration $\Delta \tau$, compute the variance and average the
results of all samples. The dotted line represents the duration
3.3 s chosen in \cite{Greinert07}. At the beginning of the
experiment, the variance of the displacement is constant for any
reasonable durations of measurement. However, we clearly see that
this method produces an underestimate of $\left<\delta x^2\right>$
for long aging times, specially when the viscoelasticity of the
gel becomes important. Long lived fluctuations are then ignored.

\subsection{Passive rheology}
\begin{figure}%[h!]
\begin{center}
\subfigure[]{
\resizebox*{6cm}{!}{\includegraphics{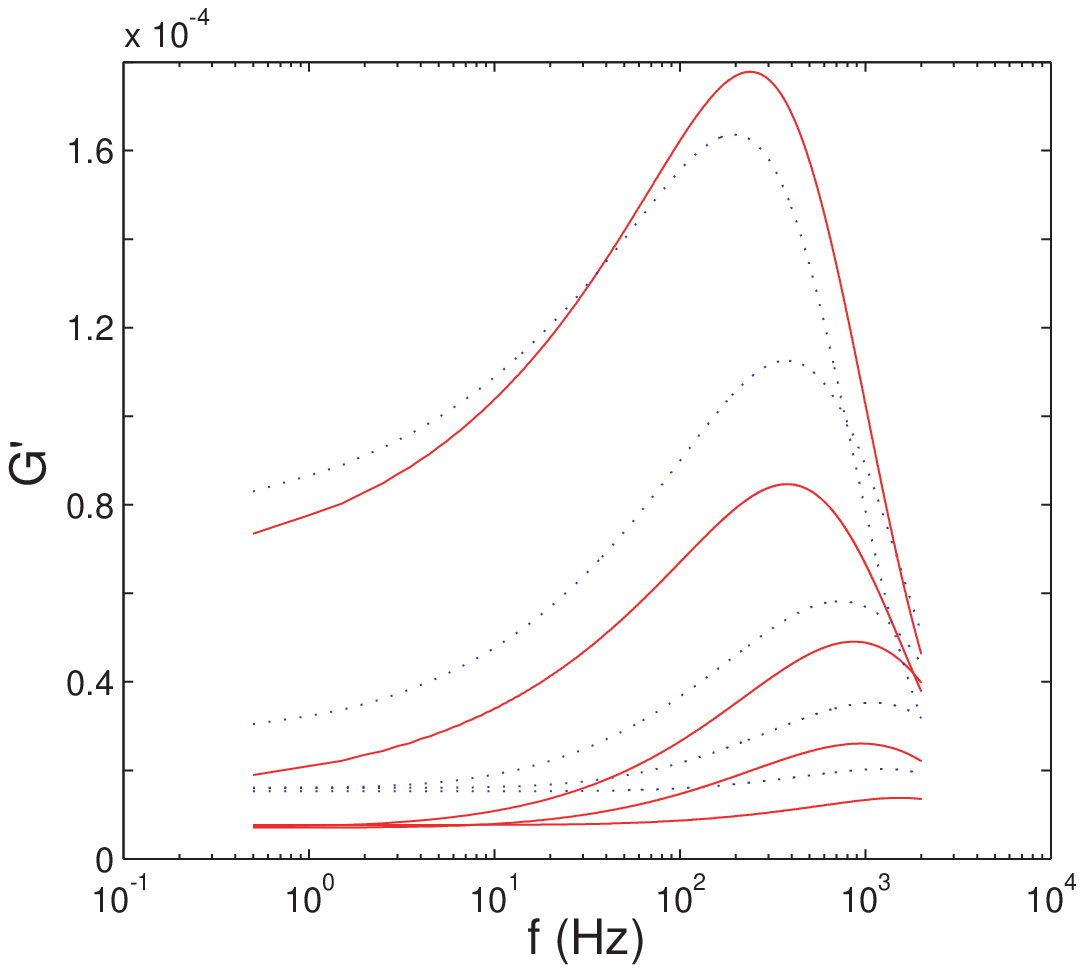}}}
\subfigure[]{
\resizebox*{6cm}{!}{\includegraphics{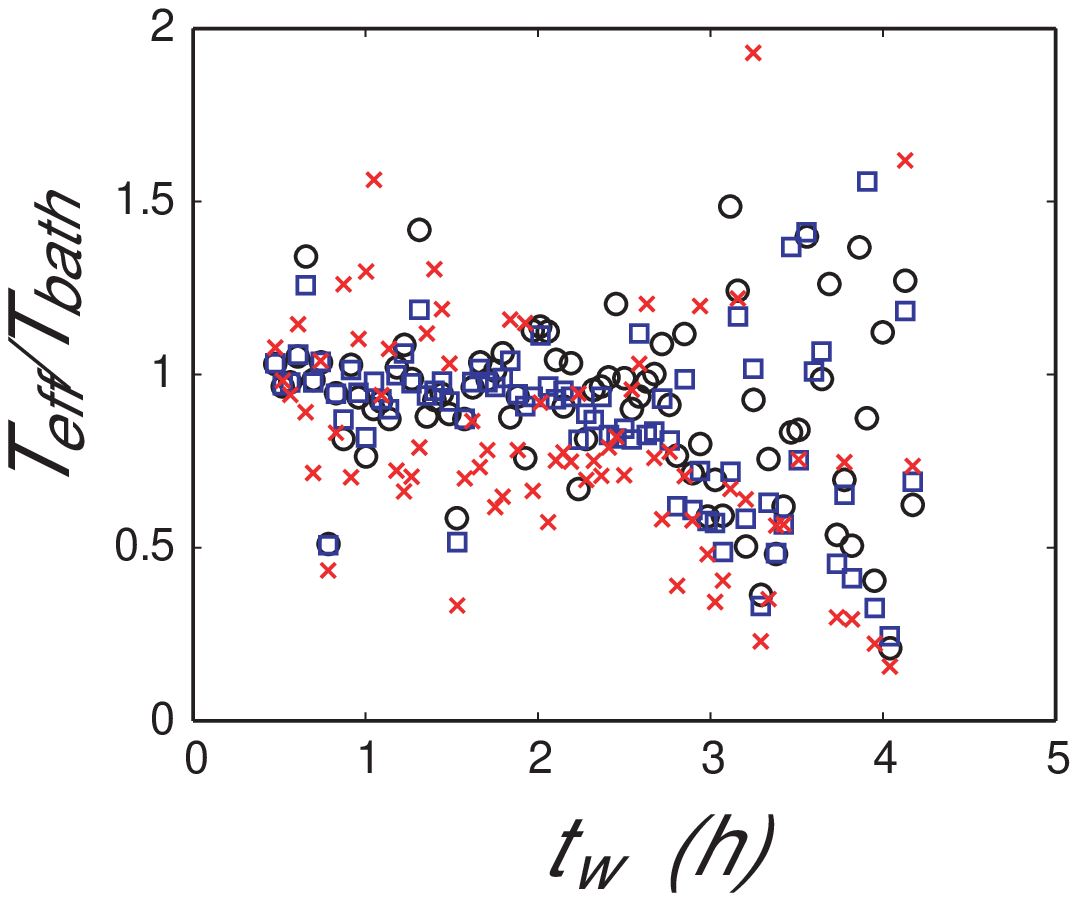}}}
\caption{\label{fig:krkr}a) Evolution of $G'$ (the global elastic modulus) as a function of frequency for both trap stiffnesses, $k_1=7.47$ pN/$\mu$m (solid lines), $k_2=16.7$ pN/$\mu$m (dashed lines) and for increasing
ageing time: $t_w=30$ min, $90$ min, $140$min, $190$ min and $240$ min (1.2\%
wt of Laponite, 2 $\mu$m glass bead and ionic strength I=5 10$^{-3}$ mol.L$^{-1}$).
  b) Evolution of the ration $T_{eff}/T_{bath}$ with ageing time ($t_w$) for different frequencies (f=1 Hz $\square$, 10 Hz $\circ$ and 100 Hz $\times$)).}
\end{center}
\end{figure}

This new method, using Kramers-Kronig relation, allows us to test
the dependence of the effective temperature on the frequency. The
figure \ref{fig:krkr}a) shows the real part of $G$, which
corresponds to the global elastic modulus of the gel and of the
laser, for both trap stiffness. This numerical method is very
sensitive to the spectrum. Thus, before computing the elastic
modulus, we average the spectrum to obtain smooth curves. The
uncertainties would then misplace the curve rather than produce a
noisy curve. The increase with time is consistent with the
increase of the strength of the gel. The elastic behavior of the
Laponite is also more pronounced at high frequency. The last
decrease of the curves at very high frequencies is due to the
numerical method. Indeed, the frequency cutoff should be set at
least a  decade below the frequency of the data acquisition: data
above 200 Hz is not reliable. We see that the curves of each
stiffness are well separated except at the end of the measurement,
where the results are not accurate due to the large difference
between the optical stiffness and the Laponite stiffness.

 From
these data, we compute the ratio of the effective temperature to
the bath temperature along the ageing process. These results are
shown on Fig. \ref{fig:krkr}b) for three different frequencies
($f=1$ Hz, 10 Hz and 100 Hz). We first note that the three curves
are almost identical. This means that the effective temperature
does not depends on the frequency. Second, the temperature ratio
is close to 1. The dispersion of the data is rather small at early
times and again increases when the stiffness of the gel overcomes
those of the optical trap. This comes from the uncertainties on
the elastic modulus which become larger than the difference
between the two curves. This dispersion may also give, for very
long time,  negative temperatures, not shown in Fig.
\ref{fig:krkr}b) which is an expanded view. Even if this method is
here less accurate than the previous one, it allows us to verify
that the effective temperature is the same for all frequencies.

\section{Conclusion}

  Finally, a remark has to be made concerning the mean position of the bead during aging. When the stiffness of
   the gel becomes comparable to the optical one, the bead starts to move away from the centre the optical trap.
   We observe a drift of the bead position at long time, which could lead to the escape of the bead.
   Moreover we have performed simultaneous measurements with a multiple trap using a fast camera showing that at very long
   $t_w$ the mean trajectories of beads separated by 7 $\mu$m are almost identical. This proves that one must pay attention
   when interpreting such measurements, specially on the duration of measurements.
We also have seen that the way the sample is sealed can accelerate
the formation of the gel and the drift of the bead by changing the
chemical properties in the small sample. We have used different
types of cell, Laponite concentrations, bead sizes,
 stiffness  of the optical trap. In each case we do not find any increase of the effective temperature.

In conclusion, our results show no increase of $T_{eff}$ in
Laponite and are in agreement with those of Jabbari-Farouji
\cite{Jabbari-Farouji07}, who measured fluctuations and responses
of the bead displacement in Laponite over a wide range of
frequency and found that $T_{eff}$ is equal to the bath
temperature.

\section{Acknowledgement} This work has been partially supported by
ANR-05-BLAN-0105-01.

%
%\begin{verbatim}
%\begin{figure}
%\begin{center}
%\subfigure[]{
%\resizebox*{5cm}{!}{\includegraphics{senu_gr1.eps}}}%
%\subfigure[]{
%\resizebox*{5cm}{!}{\includegraphics{senu_gr2.eps}}}%
%\caption{\label{fig2} Example of a two-part figure with individual %
%sub-captions showing that all lines of figure captions range left.}%
%\label{sample-figure}
%\end{center}
%\end{figure}
%\end{verbatim}

\end{document}